\documentclass[%
 aip,
 amsmath,amssymb,
 reprint,%
]{revtex4-1}

\usepackage{graphicx}
\usepackage{dcolumn}
\usepackage{bm}

\usepackage[utf8]{inputenc}
\usepackage[T1]{fontenc}
\usepackage{mathptmx}
\usepackage{etoolbox}
\usepackage{comment}

\usepackage{siunitx}
\makeatletter
\def\@email#1#2{%
 \endgroup
 \patchcmd{\titleblock@produce}
  {\frontmatter@RRAPformat}
  {\frontmatter@RRAPformat{\produce@RRAP{*#1\href{mailto:#2}{#2}}}\frontmatter@RRAPformat}
  {}{}
}%
\makeatother
\begin{document}

\preprint{AIP/123-QED}

\title{Cooperative effects of membrane confinement and gelation on PEG crystallization pathway}

\author{Masaki Yoshida}
 \thanks{These authors contributed equally to this work.}
 \affiliation{Department of Basic Science, Graduate School of Arts and Sciences, The University of Tokyo, Komaba 3-8-1, Meguro, Tokyo 153-8902, Japan}
 
 \author{Naoya Yanagisawa}%
 \thanks{These authors contributed equally to this work.}
 \affiliation{Department of Basic Science, Graduate School of Arts and Sciences, The University of Tokyo, Komaba 3-8-1, Meguro, Tokyo 153-8902, Japan}
 \affiliation{Komaba Institute for Science, Graduate School of Arts and Sciences, The University of Tokyo, Komaba 3-8-1, Meguro, Tokyo 153-8902, Japan}

\author{Fumiya Kanie}
 \thanks{These authors contributed equally to this work.}
 \affiliation{Department of Basic Science, Graduate School of Arts and Sciences, The University of Tokyo, Komaba 3-8-1, Meguro, Tokyo 153-8902, Japan}

\author{Hidemasa Teraoka}
 \affiliation{Department of Basic Science, Graduate School of Arts and Sciences, The University of Tokyo, Komaba 3-8-1, Meguro, Tokyo 153-8902, Japan}
\affiliation{Komaba Institute for Science, Graduate School of Arts and Sciences, The University of Tokyo, Komaba 3-8-1, Meguro, Tokyo 153-8902, Japan}

\author{Ikki Morichika}
 \affiliation{Institute of Industrial Science, The University of Tokyo, 4-6-1
Komaba, Meguro-ku, Tokyo 153-8505, Japan}

\author{Satoshi Ashihara}
 \affiliation{Institute of Industrial Science, The University of Tokyo, 4-6-1
Komaba, Meguro-ku, Tokyo 153-8505, Japan}

\author{Tetsuya Hama}
 \affiliation{Department of Basic Science, Graduate School of Arts and Sciences, The University of Tokyo, Komaba 3-8-1, Meguro, Tokyo 153-8902, Japan}
\affiliation{Komaba Institute for Science, Graduate School of Arts and Sciences, The University of Tokyo, Komaba 3-8-1, Meguro, Tokyo 153-8902, Japan}
 
\author{Miho Yanagisawa}
 \affiliation{Department of Basic Science, Graduate School of Arts and Sciences, The University of Tokyo, Komaba 3-8-1, Meguro, Tokyo 153-8902, Japan}
 \affiliation{Komaba Institute for Science, Graduate School of Arts and Sciences, The University of Tokyo, Komaba 3-8-1, Meguro, Tokyo 153-8902, Japan}
 \affiliation{Department of Physics, Graduate School of Science, The University of Tokyo, Hongo 7-3-1, Bunkyo, Tokyo 113-0033, Japan}
 \affiliation{Research Center for Complex Systems Biology, Universal Biology Institute, The University of Tokyo, Komaba 3-8-1, Meguro, Tokyo 153-8902, Japan}
 \email{myanagisawa@g.ecc.u-tokyo.ac.jp}

\date{\today}

\begin{abstract}
Lipid-coated microscale hydrogels provide confined, hydrated environments in which polymer phase behavior can differ markedly from that in bulk. Here, we investigate the crystallization pathway of poly(ethylene glycol) (PEG) encapsulated in lipid-coated agarose microgels. Surprisingly, polarized-light microscopy reveals birefringence in the microgels under conditions where PEG remains non-crystalline in the corresponding bulk solution. The birefringence disappears upon heating and spontaneously reappears after further cooling or upon local mechanical stimulation. Infrared microspectroscopy demonstrates that the birefringent microgels contain crystalline PEG, whereas non-birefringent microgels contain PEG in an amorphous-like state, indicating the existence of a metastable precursor prior to crystallization. Furthermore, cooling below the phase-separation temperature produces PEG-rich domains preferentially near the membrane, suggesting that membrane wetting governs the spatial distribution of PEG before crystallization. Together, these results indicate that membrane confinement and agarose gelation cooperatively alter the local hydration environment of PEG, thereby stabilizing an amorphous-like precursor and redirecting the subsequent crystallization pathway. Our findings identify the coupling of membrane wetting and gelation as a key factor governing PEG crystallization in confined soft materials.
\end{abstract}

\maketitle

\section{\label{sec:intro}Introduction}
Polymer crystallization is one of the most fundamental phase transitions in soft materials and plays a central role in determining the structure and properties of polymeric systems. Under confinement, crystallization pathways can differ substantially from those in bulk because finite dimensions, interfaces, and restricted molecular transport modify the local physicochemical environment.\cite{michell2016confined,meldrum2020crystallization,liu2021confined} Previous studies have shown that confinement influences crystal nucleation, growth kinetics, crystal morphology, molecular orientation, and polymer chain dynamics.\cite{meldrum2020crystallization,liu2021confined,shin2007enhanced, michell2013confine, michell2016confined} However, these studies have primarily focused on geometrical confinement, whereas much less attention has been paid to how confinement alters the structural state of polymers prior to crystallization and, consequently, the crystallization pathway itself.

Hydrogels provide versatile platforms for studying polymer behavior under gel-network confinement because the mesh size determines the characteristic size of the confined space. Among them, agarose is particularly attractive because it forms a stable gel network at remarkably low polymer concentrations (typically above \SIrange{0.1}{0.2}{wt\percent}).\cite{morita2013phase,ichinose2020concentration} Unlike many other hydrogel materials, agarose gelation is accompanied by phase separation at low temperatures,\cite{morita2013phase,tokita2023frictional,yanagisawa2025spatiotemporal} producing a relatively large mesh size (e.g., $\sim$1~$\mu$m at \SI{0.5}{wt\percent}) capable of accommodating macromolecules.\cite{narayanan2006determination,mao2016normal,pluen1999diffusion}
Consequently, agarose microgels have been explored as biocompatible carriers for drug delivery and other biomedical applications.\cite{wang1997preparation,yazdi2020agarose}

Poly(ethylene glycol) (PEG) is one of the most widely used hydrophilic polymers because of its excellent biocompatibility and water solubility.\cite{d2016polyethylene} The phase behavior and crystallization of PEG in bulk solutions have been extensively investigated\cite{huang2001interaction, hatakeyma2007cold}, and PEG crystallization has also been studied in PEG/agarose composite materials and PEG-based hydrogels.\cite{csenturk2011biodeg, mishra2020PEG, zhang2018influence} However, how confinement within a hydrated microscale gel network affects the crystallization pathway of dilute PEG remains largely unexplored.

Membrane-coated agarose microgels provide a unique system in which gel-network confinement is coupled with membrane confinement. Previous studies have shown that membrane wetting modifies the spatial distribution of encapsulated polymers \cite{kanakubo2023cell,hester2023fluid,cao2015one} and can induce liquid--liquid phase separation in membrane-confined polymer droplets and microgels\cite{watanabe2022cell,yanagisawa2014multiple,yanagisawa2022cell}.
Interfacial interactions can also modify polymer crystallization by altering the thermodynamic and kinetic conditions for nucleation.\cite{dolynchuk2023thermodynamics} However, it remains unknown whether these coupled effects alter the crystallization pathway of encapsulated polymers.

In this study, we investigate the crystallization pathway of PEG encapsulated in phospholipid-coated agarose microgels. Using polarized-light microscopy and infrared microspectroscopy, we show that PEG crystallizes under conditions where it remains non-crystalline in the corresponding bulk solution by first forming a metastable amorphous-like precursor. We further show that membrane wetting governs the spatial distribution of PEG before crystallization. These observations support a mechanism in which gelation changes the local hydration state of PEG while membrane wetting biases its spatial distribution, with combined effects favoring the formation of a metastable precursor before crystallization. This work reveals how the coupling between a gel network and a membrane interface can redirect polymer crystallization away from its bulk pathway.

\begin{figure*}[htbp]
    \centering
    \includegraphics[width=0.95\textwidth]{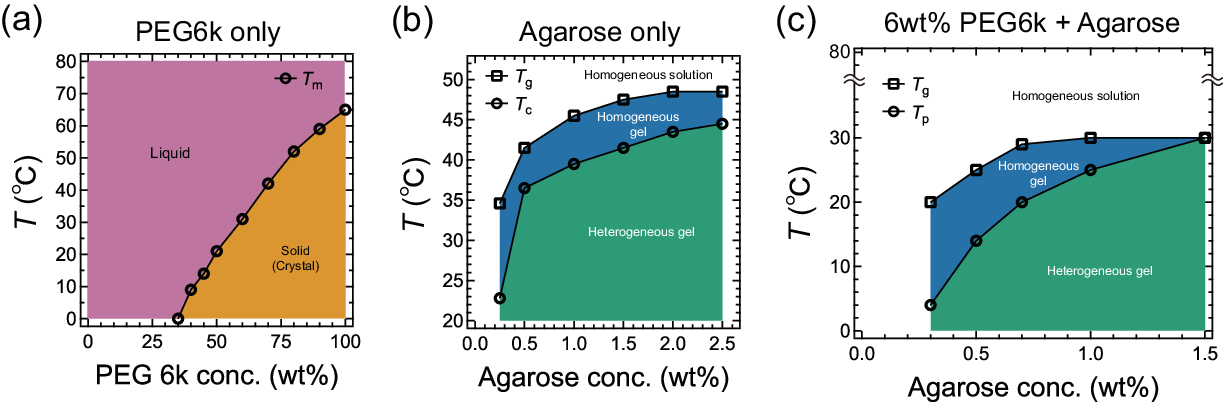}
    \caption{Phase diagrams of (a) PEG6k solutions, (b) agarose solutions, and (c) PEG6k/agarose mixtures. (a) The melting temperature ($T_{\mathrm{m}}$), separating the liquid and solid phases of PEG6k, was determined by polarized light microscopy. (b) The gelation temperature ($T_{\mathrm{g}}$), separating the liquid and gel phases, and the transition temperature ($T_{\mathrm{c}}$), separating homogeneous and heterogeneous gel phases, were reproduced from Ref.\cite{morita2013phase} with permission. $T_{\mathrm{g}}$ and $T_{\mathrm{c}}$ were determined by the tilting method and light scattering, respectively. (c) The gelation temperature ($T_{\mathrm{g}}$), separating the liquid and gel phases, and the phase-separation temperature ($T_{\mathrm{p}}$), separating homogeneous and heterogeneous phases, were determined by the tilting method and microscopic observation, respectively.}
    \label{fig:fig1}
\end{figure*}

\section{\label{sec:method}Experimental}
\subsection{Materials}
Agarose (type IV, Lot No. 043K1302) was purchased from Sigma-Aldrich (St. Louis, MO, USA). Its nominal weight-average molecular weight ($M_{\mathrm{w}}$), determined by gel permeation chromatography, was 2.9 $\times$ 10$^{2}$ kg/mol. 
We used distilled water (Catalog No. 10977-023, Invitrogen, Carlsbad, CA, USA) as the solvent. 
Poly(ethylene glycol) with a weight-average molecular weight of 7--9 kg/mol (PEG6k; Catalog No. 169-09125, FUJIFILM Wako Pure Chemical Corporation, Osaka, Japan) was used. Mineral oil (Nacalai Tesque, Kyoto, Japan) and hexadecane (Tokyo Chemical Industry Co., Ltd., Tokyo, Japan) were used as the oil phase for preparing agarose droplets. 
The electrically neutral phospholipid 1-palmitoyl-2-oleoyl-sn-glycero-3-phosphocholine (PC; Avanti Polar Lipids, Alabaster, AL, USA) was used to stabilize the agarose droplets by forming a lipid monolayer at the oil--water interface. PC was dissolved in chloroform (FUJIFILM Wako Pure Chemical Corporation, Osaka, Japan) before being added to the oil phase. 
To visualize the PEG-rich phase, rhodamine B-labeled methyl poly(ethylene glycol) ($M_{\mathrm{w}}$ = 5 kg/mol; RB-PEG, Catalog No. FL001045-5k, Biopharma PEG Scientific Inc., Watertown, MA, USA) was added. All chemicals were used as received without further purification.

\subsection{\label{sec:method_sample}Preparation of microgels and bulk samples}
We first prepared a 10 mM PC solution in chloroform. 
Fifty microliters of the PC solution were added to 500 $\mu$L of mineral oil or hexadecane in a vial. 
The vial was covered with sealing tape perforated with several small holes and kept at 70$^\circ$C for 24 h to evaporate the chloroform. 
The oil phase was then vortex-mixed thoroughly. 
An aqueous solution containing 6 wt\% PEG and 0.5 wt\% agarose was prepared. 
The solution was dissolved by repeating three cycles of vortex mixing (1--2 min) and heating at 80$^\circ$C (1--2 min) and then allowed to stand for 1 hour. 
Next, 6 $\mu$L of the PEG--agarose solution was added to 100 $\mu$L of the PC-containing oil phase and emulsified by pipetting to prepare PC-coated droplets. 
During emulsification, the sample was maintained at 80$^\circ$C to prevent phase separation. 
As control samples, droplets containing only agarose or only PEG were prepared in the same manner. Immediately after emulsification, the droplets were transferred to a glass-bottom dish (35-mm dish with a 14-mm glass coverslip; Matsunami Glass Ind., Ltd., Osaka, Japan) and kept at room temperature (approximately 25 $^\circ$C) for 1 h to allow the agarose to gel.

\subsection{Microscopic observation}
Microgels in glass-bottom dishes were observed using a polarizing microscope (BX53, Olympus, Tokyo, Japan) equipped with a high-speed camera (HAS-D71, Ditect, Tokyo, Japan; 8,000 frames s$^{-1}$). With crossed polarizers, we mechanically stimulated the microgels by gently contacting their surfaces with a gel-loading pipette tip and observed the resulting birefringence. The distribution of RB-PEG ($\sim$1\% of PEG) within the microgels was examined using a confocal laser scanning microscope (IX83/FV1200, Olympus, Tokyo, Japan) with excitation at 473 nm and fluorescence collected between 490 and 590 nm.

\subsection{\label{sec:method_micro}Infrared microspectroscopy}
Infrared spectra of the microgels were acquired using a Fourier transform infrared microscope (Nicolet iN10, Thermo Fisher Scientific, Waltham, MA, USA) operated in transmission mode. The measurements were performed with a spectral resolution of 4~cm$^{-1}$ using an MCT detector, and 128 scans were accumulated for each spectrum. Approximately 3 $\mu$L of oil containing dispersed microgels was placed on a ZnSe substrate (13 mm diameter, 1 mm thickness; Pier Optics Co., Ltd., Gunma, Japan). The surrounding oil phase was used as the background, and the spectrum of an individual microgel was recorded. The infrared absorption spectrum of PEG was obtained by subtracting the contributions of the oil phase and agarose from the measured spectrum.

PEG/agarose microgels were prepared as described above (see Experimental section \ref{sec:method_sample}). Bulk PEG solutions were prepared by dissolving PEG in distilled water at the desired concentrations. To obtain reference spectra of amorphous and crystalline PEG, PEG solutions were allowed to evaporate in glass-bottom dishes at room temperature. Samples that exhibited no birefringence under crossed polarizers were used as amorphous references, whereas samples exhibiting birefringence were used as crystalline references.

\subsection{Interfacial tension measurement}
The interfacial tension between aqueous PEG or agarose droplets and the surrounding oil phase was measured at approximately 40~$^\circ$C using the pendant drop method (DM-501, Kyowa Interface Science Co., Ltd., Saitama, Japan). Mineral oil containing PC at a concentration of approximately 1~mg/mL was used as the oil phase. The measurements and analysis were performed following the procedure described previously.\cite{kanakubo2023cell} For droplets containing 0.5~wt\% agarose solution, direct determination of the interfacial tension was difficult because of the high viscosity of the solution. Therefore, the time-dependent relaxation of the interfacial tension was analyzed by fitting the relaxation curve with a single-exponential function proportional to $\exp(-t/\tau)$, where $\tau$ denotes the characteristic relaxation time. The interfacial tension at $t=\tau$ was used as a representative value for comparison with that of the PEG solution.

\section{\label{result}Results and Discussion}

\subsection{Birefringence of PC-coated agarose/PEG microgels}
Figure 1 summarizes the phase diagrams of PEG6k, agarose, and PEG6k/agarose mixtures. Pure PEG6k solutions crystallize below the melting temperature ($T_{\mathrm{m}}$) (Fig. 1a).\cite{huang2001interaction} For example, $T_{\mathrm{m}}$ exceeds 0 $^\circ$C at PEG concentrations above approximately 25 wt\%. In contrast, agarose solutions undergo gelation below the gelation temperature ($T_{\mathrm{g}}$) (Fig. 1b). For 0.5 wt\% agarose, $T_{\mathrm{g}}$ is approximately 40 $^\circ$C. Upon deeper quenching below $T_{\mathrm{g}}$, the gel becomes opaque owing to the coexistence of agarose-rich and agarose-poor phases, as reported previously.\cite{morita2013phase,yanagisawa2025spatiotemporal} Figure 1c shows the phase diagram of PEG6k/agarose mixtures containing 6 wt\% PEG6k and various concentrations of agarose. For the mixture containing 6 wt\% PEG6k and 0.5 wt\% agarose, gelation occurs at approximately 26 $^\circ$C, whereas phase separation occurs only below approximately 14 $^\circ$C. Therefore, quenching the homogeneous solution from 80 to 25 $^\circ$C induces gelation while avoiding both PEG crystallization and phase separation. Unless otherwise noted, the following experiments were performed using this composition and temperature protocol.
\begin{figure*}[htbp]
    \centering
    \includegraphics[width=0.95\textwidth]{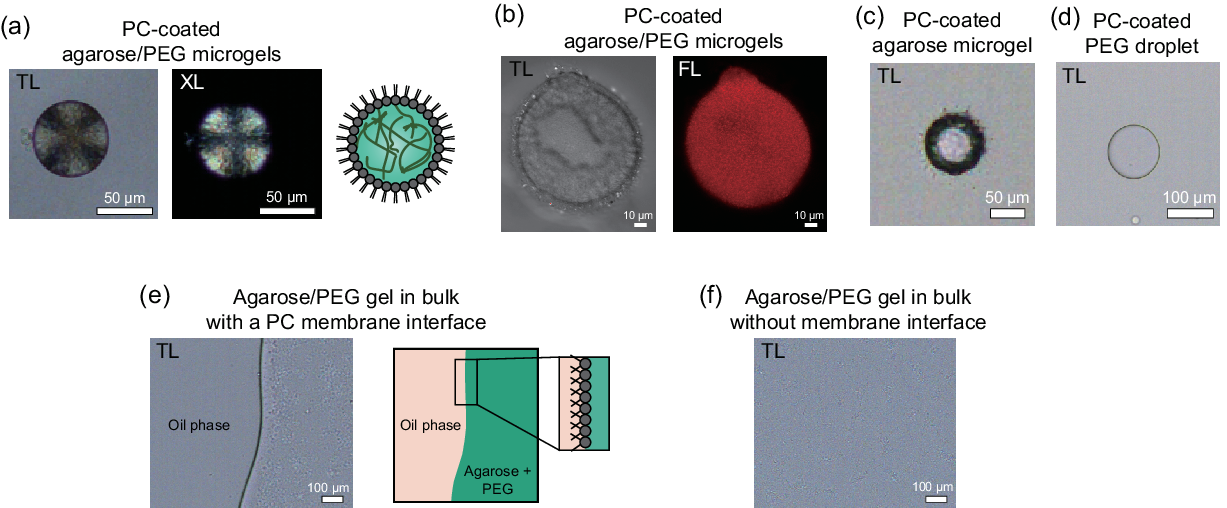}
    \caption{(a) Transmitted-light (TL) and crossed-polarizer (XL) images of a PC-coated agarose/PEG microgel. (b) TL and fluorescence images of a PC-coated agarose/PEG microgel containing rhodamine B-labeled PEG (RB-PEG; shown in red). (c--f) TL images of (c) a PC-coated agarose microgel, (d) a PC-coated PEG solution droplet, and bulk agarose/PEG gels with (e) and without (f) a PC monolayer at the oil/water interface. The PEG6k and agarose concentrations were 6 wt\% and 0.5 wt\%, respectively.}
    \label{fig:fig2}
\end{figure*}
Following the phase behavior shown in Fig. 1, microgels containing 6 wt\% PEG6k and 0.5 wt\% agarose was prepared by emulsifying the homogeneous solution at 80 $^\circ$C and subsequently quenching it to 25 $^\circ$C (see Experimental section \ref{sec:method_sample}\ref{sec:method_micro}). Under these conditions, the bulk phase diagram predicts gelation without either PEG crystallization or macroscopic phase separation. Nevertheless, polarized light microscopy revealed that, although the PC-coated agarose/PEG microgels appeared uniform in transmitted-light (TL) images, they exhibited distinct birefringence patterns under crossed polarizers (XL) (Fig. 2a). Thus, birefringence emerged under conditions where PEG crystallization is not expected.

We first examined whether the birefringence was associated with microscale phase separation by replacing a fraction of the PEG with rhodamine B-labeled PEG (RB-PEG). Confocal fluorescence microscopy showed a homogeneous fluorescence distribution throughout the birefringent microgels, indicating that PEG-rich domains larger than the spatial resolution of the microscope of $\sim$1 $\mu$m were absent (Fig. 2b).

We next examined the conditions required for birefringence using several control samples (Fig. 2c--f). No birefringence was observed in PC-coated agarose microgels (Fig. 2c) or in PC-coated PEG solution droplets (Fig. 2d), indicating that neither agarose nor PEG alone is sufficient to generate birefringence. We next examined bulk agarose/PEG gels to evaluate the role of the membrane interface. No birefringence was detected, even in bulk gels partially in contact with a PC monolayer at the oil/water interface (Fig. 2e), and the same result was obtained in the absence of the PC monolayer (Fig. 2f). These results indicate that the coexistence of PEG and a gel network is not sufficient to produce birefringence. Instead, birefringence is observed only when the polymer mixture is confined within membrane-coated microgels.

We further examined whether this phenomenon was specific to agarose by preparing PC-coated gelatin/PEG microgels under conditions where gelatin formed a gel (1.7 wt\% PEG6k and 5 wt\% gelatin as reported in our previous paper\cite{yanagisawa2014multiple}). We also observed birefringence in these microgels (data not shown), suggesting that the phenomenon is not specific to the chemical nature of agarose but may instead arise from the combination of gelation and membrane confinement.

\begin{figure*}[htbp]
    \centering
    \includegraphics[width=0.9\textwidth]{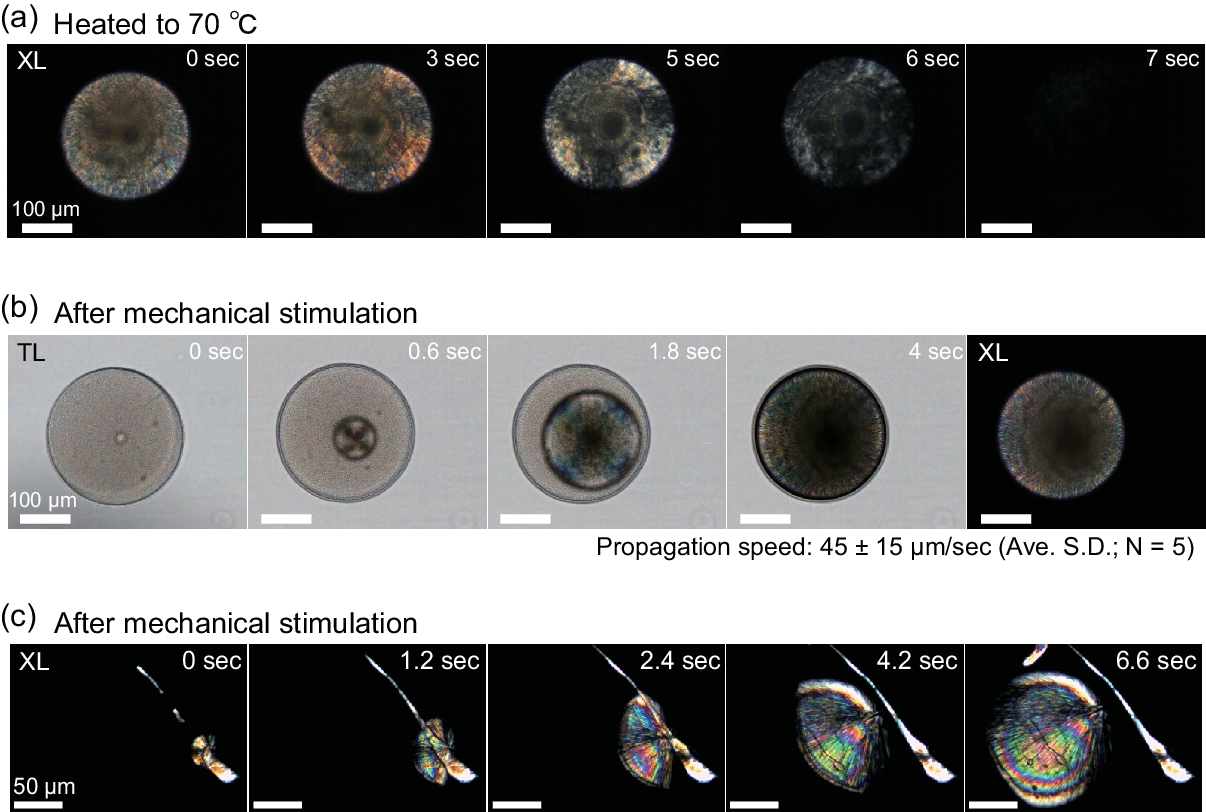}
    \caption{(a) Crossed-polarizer (XL) images of a PC-coated agarose/PEG microgel before and after heating to 70 $^\circ$C. The corresponding time evolution is shown in Supplementary Movie~S1. The birefringence disappeared upon heating. (b, c) Transmitted-light (TL) and crossed-polarizer (XL) images before and after local mechanical stimulation of a PC-coated agarose/PEG microgel by gentle contact with a gel-loading pipette tip, showing the reappearance of birefringence (Supplementary Movies~S2 and S3).}
    \label{fig:fig3}
\end{figure*}

\begin{figure*}[htbp]
    \centering
    \includegraphics[width=0.95\textwidth]{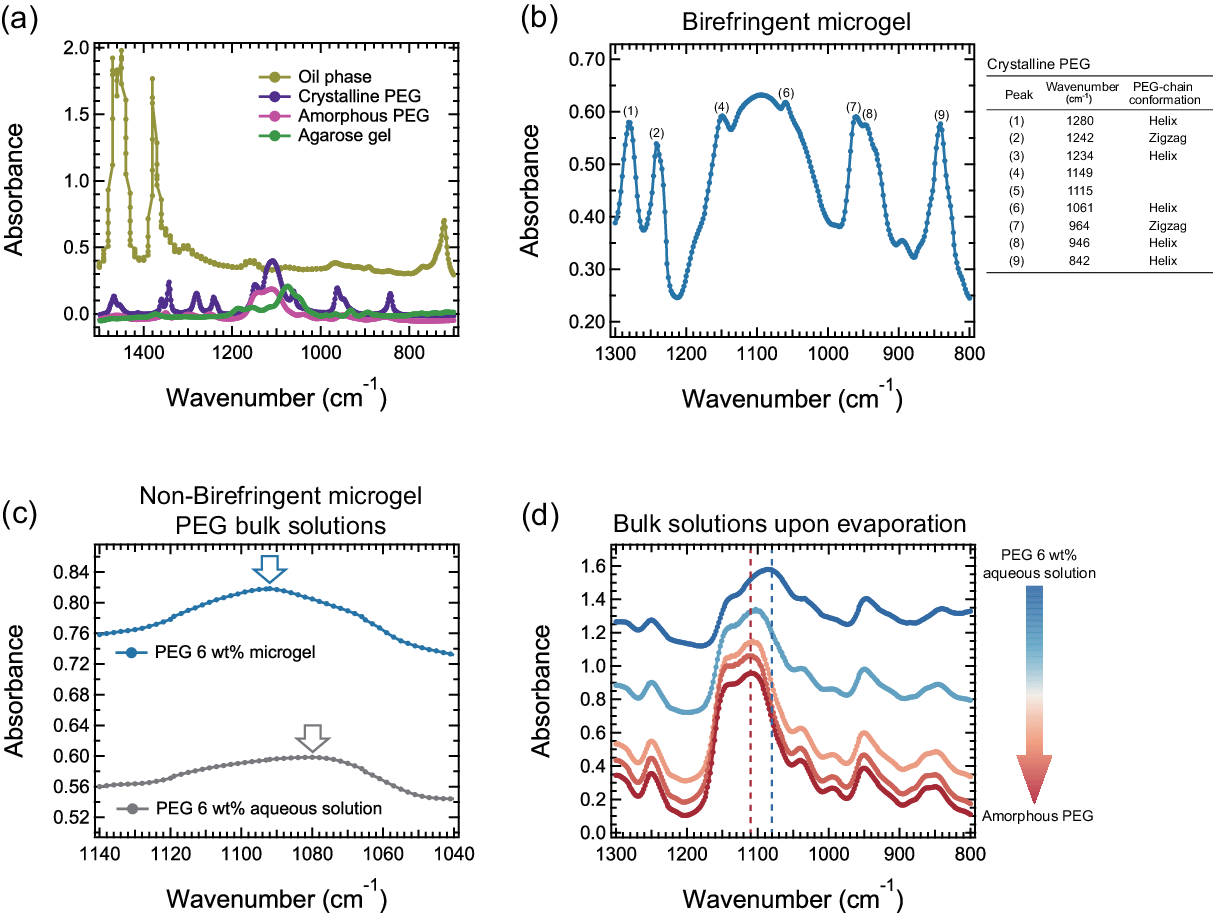}
    \caption{(a) Reference FTIR spectra of the oil phase, agarose gel, amorphous-like PEG, and crystalline PEG. (b) FTIR spectrum of a birefringent PC-coated agarose/PEG microgel after subtraction of the contributions from the oil phase and agarose. The seven characteristic peaks agree with those of crystalline PEG reported previously.~\cite{youjun2006bonding} (c) Comparison of the FTIR spectra of PEG in non-birefringent PC-coated agarose/PEG microgels and 6 wt\% PEG solution. Unlike the PEG bulk solutions, which have a peak at approximately 1080 cm$^{-1}$ (gray) regardless of concentration (up to 30 wt\%, not shown), the non-birefringent microgel exhibits a peak at a higher position, approximately 1090$\sim$1100 cm$^{-1}$ (blue). (d) Change in FTIR spectra of PEG solution with water evaporation. The peak at approximately 1080 cm$^{-1}$ (shown in blue) shifts to a higher position, closer to approximately 1111 cm$^{-1}$ (red), suggesting the PEG transition from the liquid phase to an amorphous-like state.}
    \label{fig:fig4}
\end{figure*}
To examine the relationship between birefringence and the gel state, the microgels were heated to 70 $^\circ$C, above the melting temperatures of both PEG6k crystals and the agarose gel (Fig. 1a, b). As shown in Fig. 3a and Supplementary Movie~S1, the birefringence disappeared completely upon heating. We next examined the response of the microgels to mechanical stimulation. Gentle contact of the microgel surface with a gel-loading pipette tip triggered the reappearance of birefringence near the contact region (Figs. 3b and 3c; Supplementary Movies~S2 and S3). The birefringence propagated through the microgel with a velocity of 45 $\pm$ 15 $\mu$m s$^{-1}$ (mean $\pm$ S.D., N = 5), indicating that local mechanical stimulation initiated a structural transformation that propagated through the microgel rather than producing an instantaneous elastic response.
\begin{figure*}[htbp]
    \centering
    \includegraphics[width=0.95\textwidth]{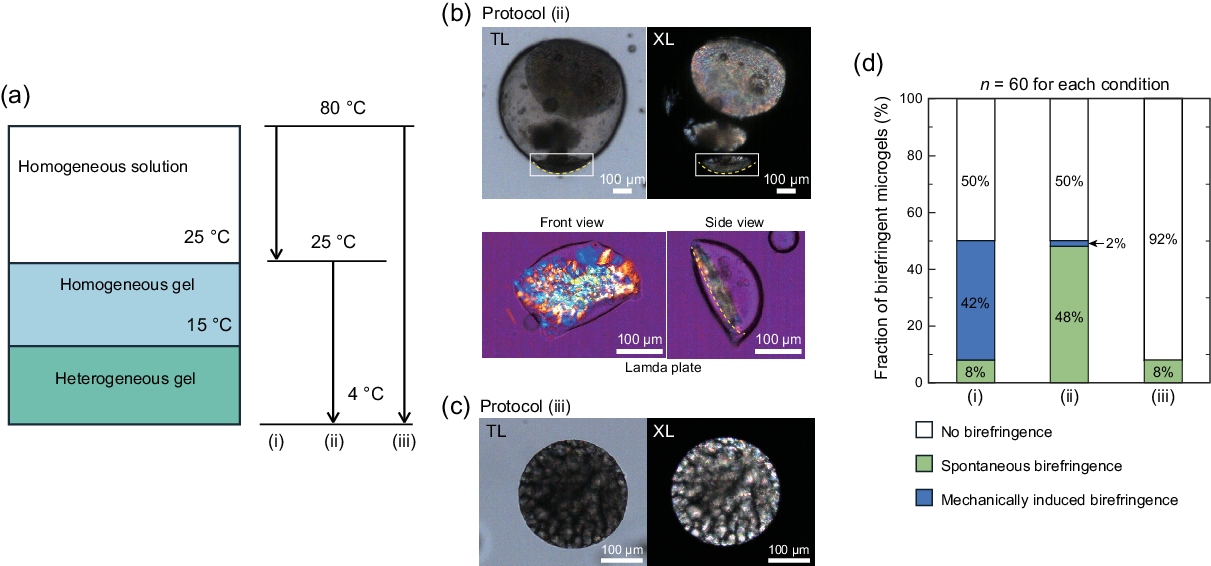}
    \caption{(a) Temperature protocols used to investigate PEG crystallization in PC-coated agarose/PEG microgels: (i) cooling from 80 to 25 $^\circ$C; (ii) cooling from 80 to 25 $^\circ$C, holding at 25 $^\circ$C for approximately 1 h, and subsequently cooling to 4 $^\circ$C; and (iii) rapid cooling from 80 to 4 $^\circ$C, passing through the phase-separation temperature (15 $^\circ$C). (b) Transmitted-light (TL) and crossed-polarizer (XL) images of a microgel prepared using protocol (ii). Dark aggregates in the TL image correspond to PEG crystals. The enlarged image obtained with a first-order retardation ($\lambda$) plate shows that PEG crystals are preferentially located near the membrane surrounding the microgel. (c) TL and XL images of a microgel prepared using protocol (iii). Compared with protocol (ii), we observe more, smaller birefringent domains. (d) Fractions of microgels exhibiting no birefringence, spontaneous birefringence, or mechanically induced birefringence under each temperature protocol ($n = 60$ microgels per condition).}
    \label{fig:fig5}
\end{figure*}
\subsection{Infrared microspectroscopy of agarose/PEG microgels}
To identify the structural origin of the birefringence, we performed infrared microspectroscopy on the PC-coated agarose/PEG microgels (see Experimental Sections \ref{sec:method_sample} for sample preparation and measurement). First, reference spectra were obtained for the individual components of the system, including the oil phase, agarose gel, and PEG in its amorphous-like and crystalline states (Fig. \ref{fig:fig4}a). Comparison of the amorphous and crystalline PEG spectra revealed pronounced differences in the 800--1300 cm$^{-1}$ region. Therefore, subsequent analyses focused on this spectral region.

Figure \ref{fig:fig4}b shows the infrared spectrum of a birefringent microgel. After subtracting the contributions of the oil phase and agarose, the resulting PEG spectrum exhibited seven characteristic absorption bands of crystalline PEG reported previously.~\cite{youjun2006bonding} These results strongly suggest that the birefringence observed in the PC-coated agarose/PEG microgels originates from PEG crystallization.

We next investigated the structural state of PEG before birefringence appears. Figure \ref{fig:fig4}c compares the spectra of PEG in non-birefringent microgels with those of aqueous PEG solutions. In aqueous PEG bulk solutions, the absorption band near 1080 cm$^{-1}$ remained essentially unchanged over the concentration range examined (Fig. \ref{fig:fig4}c). In contrast, gradual removal of water by solvent evaporation caused this band to shift continuously from 1080 cm$^{-1}$ toward higher wavenumbers, reaching approximately 1111 cm$^{-1}$, suggesting the transition from the aqueous PEG state to an amorphous-like state (Fig. \ref{fig:fig4}d). Interestingly, the PEG spectrum of non-birefringent agarose/PEG microgels exhibited a broad absorption peak at a similar position, approximately 1090$\sim$1100 cm$^{-1}$ (Fig. \ref{fig:fig4}c). Upon the appearance of birefringence, the spectrum changed further and became identical to that of crystalline PEG (Fig. \ref{fig:fig4}a). These results suggest that, within the membrane-confined microgels, PEG first transitions to an amorphous-like state and subsequently crystallizes, giving rise to the observed birefringence.

\subsection{\label{sec:method_temp}Temperature protocols}
To investigate the effect of phase separation on the formation of birefringence, three temperature protocols were examined (Fig. 5a). Under protocol (ii), the microgels were first cooled to 25 $^\circ$C and held for approximately 1 h before further cooling to 4 $^\circ$C, thereby crossing the phase-separation temperature under near-equilibrium conditions.

As shown in Fig. 5b, numerous dark domains appeared in the transmitted-light (TL) images after cooling to 4 $^\circ$C. Observation with a first-order retardation ($\lambda$) plate revealed that these domains were preferentially located near the membrane covering the microgel. This localization is consistent with the preferential wetting of PEG at lipid membranes, as PEG has a lower interfacial tension against the PC membrane than agarose (The interfacial tension is 22.1 $\pm$ 1.4 and 17.3 $\pm$ 0.6 mN/m (Ave. $\pm$ S.D.; n = 5) for agarose and PEG solutions, respectively).

In contrast to protocol (i), under which birefringence often appeared after mechanical stimulation, microgels prepared using protocol (ii) frequently exhibited spontaneous birefringence (Fig. 5d), indicating that the ordered state can develop without external mechanical perturbation under these conditions.

Under protocol (iii), in which the microgels were quenched directly from 80 to 4 $^\circ$C, the number of dark domains increased markedly, whereas their size became considerably smaller than those observed under protocol (ii) (Fig. 5c). Despite the increased number of domains, spontaneous birefringence was rarely observed (Fig. 5d), suggesting that rapid quenching promotes the formation of numerous small PEG-rich domains while reducing the probability of birefringence.

\begin{figure*}[htbp]
    \centering
    \includegraphics[width=0.85\textwidth]{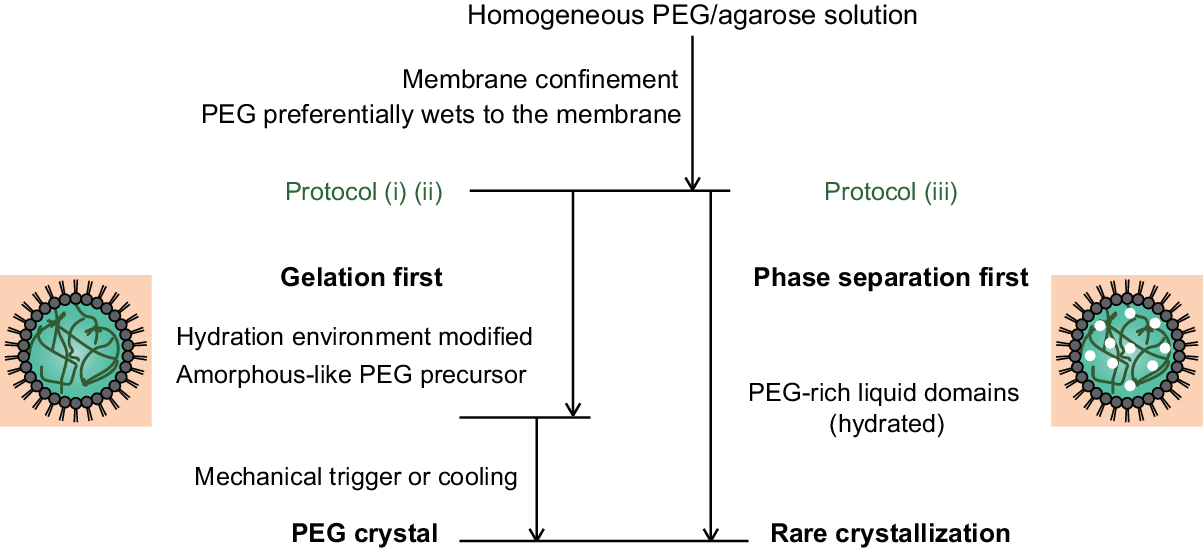}
    \caption{Proposed mechanism for crystallization of PEG inside PC-coated microgels.}
    \label{fig:fig6}
\end{figure*}

\section{Discussion}
Based on experimental observations, we propose the mechanism shown in Fig. 6. The crystallization behavior of PEG is governed by the interplay among membrane wetting, agarose gelation, and thermal history. Under protocol (iii), rapid quenching from 80 to 4 $^\circ$C causes the system to cross the phase-separation temperature before agarose gelation is fully established. As a result, PEG-rich liquid domains are formed within the microgel. Because PEG preferentially wets the membrane interface, these PEG-rich domains localize near the membrane. However, because they remain highly hydrated, only a small fraction subsequently develops birefringence.

In contrast, under protocols (i) and (ii), agarose gelation precedes without phase separation. We propose that membrane wetting enriches PEG near the membrane, while gelation modifies the hydration environment of PEG by reducing the water available for PEG hydration, as suggested previously~\cite{xiong2005topology, hatakeyma2007cold}. Together, these effects stabilize a highly concentrated amorphous-like PEG state. This interpretation is supported by the FTIR spectra, in which the non-birefringent microgels exhibit the same absorption band near 1110 cm$^{-1}$ as amorphous PEG prepared by solvent evaporation (Fig. 4c, d).

Once this amorphous-like precursor state is established, PEG readily crystallizes either by local mechanical stimulation (Fig. 3a, b) or by further cooling (Fig. 5c). Mechanical stimulation provides local crystal nucleation, followed by crystal growth throughout the microgel, whereas further cooling promotes spontaneous nucleation and growth without external perturbation.

This model consistently explains the dependence of birefringence on thermal history, the preferential localization of birefringent domains near the membrane (Fig.~5b), the mechanically triggered emergence and propagation of birefringence (Fig.~3b, c), and the FTIR evidence for an amorphous-like precursor preceding PEG crystallization. The existence of such a precursor is consistent with previously reported two-step crystallization pathways in polymers, in which disordered or amorphous-like states precede the formation of crystalline structures~\cite{ma2017tuning}.
To examine whether the observed phenomenon is specific to agarose, we also prepared PC-coated gelatin/PEG microgels and observed similar birefringence. Although infrared spectroscopy did not directly characterize the structural state of PEG in these microgels, this result supports the possibility that the phenomenon is governed by a combination of gelation and membrane confinement rather than being specific to agarose.
These results provide a broader perspective in which membrane confinement and gelation can control polymer crystallization pathways by stabilizing a metastable precursor state.

\section{Summary}
We demonstrated that membrane-coated agarose microgels provide an environment in which PEG crystallizes under conditions where crystallization does not occur in the corresponding bulk solution. Prior to crystallization, PEG forms a non-birefringent amorphous-like precursor, whereas its subsequent crystallization can be initiated either mechanically or thermally. Our results further indicate that membrane wetting controls the spatial distribution of PEG, while gelation modifies its local hydration environment. The coupling of these effects therefore provides a route by which membrane and gel-network confinement can redirect polymer crystallization pathways.

\begin{acknowledgments}
This research was partially funded by the Japan Society for the Promotion of Science (JSPS) KAKENHI (grant nos. 22H01188, 24H02287) (M. Y.), the Japan Science and Technology Agency (JST) (grant nos. FOREST, JPMJFR213Y; CREST (JPMJCR22E1)) (M. Y.). We thank Prof. Hajime Tanaka (The University of Tokyo) for the insightful suggestion, early in this work, that the observed birefringence could originate from PEG crystallization. We also thank Keisuke Koyanagi for his assistance in constructing the phase diagrams.
\end{acknowledgments}

\section*{Data Availability Statement}
The data that support the findings of this study are available from the corresponding author upon reasonable request.



\bibliography{aipsamp}

\end{document}